\documentclass[aps,prl,twocolumn,showpacs,superscriptaddress]{revtex4}
\usepackage{amssymb,graphicx,amsmath,enumerate}

\begin{document}

\title{Evidence for localization and 0.7 anomaly in hole quantum point contacts}

\author{Y. Komijani}
\author{M. Csontos}
\affiliation{Solid State Physics Laboratory, ETH Zurich, 8093 Zurich, Switzerland}

\author{I. Shorubalko}
\affiliation{Solid State Physics Laboratory, ETH Zurich, 8093 Zurich, Switzerland}
\affiliation{Electronics/Metrology/Reliability Laboratory, EMPA, 8600 Duebendorf, Switzerland}

\author{T. Ihn}
\author{K. Ensslin}
\affiliation{Solid State Physics Laboratory, ETH Zurich, 8093 Zurich, Switzerland}

\author{Y. Meir}
\affiliation{Physics Department, Ben Gurion University, Beer Sheva 84105, Israel}

\author{D. Reuter}
\author{A. D. Wieck}
\affiliation{Angewandte Festk\"orperphysik, Ruhr-Universit\"at Bochum, 44780 Bochum, Germany}

\date{\today}

\begin{abstract}

Quantum point contacts implemented in p-type GaAs/AlGaAs heterostructures are investigated by low-temperature electrical conductance spectroscopy measurements. Besides one-dimensional conductance quantization in units of $2e^{2}/h$ a pronounced extra plateau is found at about $0.7(2e^{2}/h)$ which possesses the characteristic properties of the so-called ``0.7 anomaly'' known from experiments with n-type samples. The evolution of the 0.7 plateau in high perpendicular magnetic field reveals the existence of a quasi-localized state and supports the explanation of the 0.7 anomaly based on self-consistent charge localization. These observations are robust when lateral electrical fields are applied which shift the relative position of the electron wavefunction in the quantum point contact, testifying to the intrinsic nature of the underlying physics.

\end{abstract}


\maketitle

Since its discovery in 1988 \cite{Wees1988,Wharam1988} conductance quantization in units of 2$e^2/h$ in ballistic quantum point contacts (QPCs) has been studied for various QPC geometries \cite{Buttiker1990}. In addition to the conductance plateaus at integer multiples of 2$e^2/h$, in most QPC geometries an extra plateau arises around 0.7$(2e^2/h)$ \cite{Thomas1996}. This feature evolves smoothly into the spin-resolved $e^{2}/h$ plateau at high \emph{in-plane} magnetic fields \cite{Thomas1996} revealing its spin-related nature. Possible explanations to date have focused on many-body phenomena, such as spontaneous static spin-polarization \cite{Kristensen2000,Berggren2002,Reilly2002,Reilly2005,Rokhinson2006,Sfigakis2008}, separation of singlet and triplet channels \cite{Rejec2000}, or spin and charge channels \cite{Matveev2004}. An alternative explanation suggests that the conductance is suppressed due to Coulomb repulsion from a quasi-localized state in the QPC \cite{Meir2002,Meir2008}, and restored to the 2$e^2/h$ full value at low temperatures due to the Kondo effect. The emergence of such a quasi-localized state in QPCs has been predicted based on spin-density functional theory (SDFT) calculations \cite{Hirose2003,Rejec2006}, and quantum Monte Carlo calculations \cite{Guclu2009}. Kondo-like physics has indeed been observed in QPCs \cite{Cronenwett2002}.

The more pronounced carrier-carrier interactions in low-dimensional hole systems \cite{Grbic2007,Komijani2008} compared to their n-type counterparts make p-doped systems especially suitable for investigating many-body effects such as the 0.7 anomaly \cite{Hamilton2008}. While previous studies \cite{Rokhinson2006,Hamilton2008,Danneau2008,Klochan2006,Shabani2008} focused on Si-doped (311) structures with an anisotropic in-plane Fermi surface, we present data on C-doped (100) samples. We performed conductance spectroscopy measurements of hole QPCs oriented along the high symmetry crystallographic directions parallel to the cleaved edges of the wafer at magnetic fields applied \emph{perpendicular} to the plane of the two-dimensional hole gas (2DHG). We found that the 0.7 anomaly gradually evolves into a Coulomb resonance-like peak at high magnetic fields accompanied by a Coulomb blockade diamond observed in the finite bias conductivity. In symmetrically designed QPCs both features are insensitive to a lateral displacement of the wavefunction in the QPC channel. This provides experimental evidence for the intrinsic origin of the quasi-localized state.

Measurements were performed on five QPCs with different geometries, all based on the same wafer material. The host heterostructure consists of a 5 nm undoped GaAs cap layer, followed by a 15 nm thick, homogeneously C-doped layer of AlGaAs separated from the 2DHG formed in the electronically isotropic (100) plane by a 25 nm thick, undoped AlGaAs spacer layer \cite{Reuter1999}. Prior to sample fabrication the quality of the 2DHG ($n$ = 4$\times$10$^{11}$ cm$^{-2}$, $\mu$ = 120'000 cm$^{2}$/Vs) was characterized by standard magnetotransport measurements at 4.2 K \cite{Grbic2004}. Typical values for the interaction parameter $r_{s} = E_{\rm int}/E_{\rm F}$ are $r_{s}>5$.

Investigations of the 0.7 anomaly in hole systems are experimentally challenging, because the disordered background potential can give rise to conductance resonances visible in the linear conductance at the rise between pinch-off and the first quantized plateau \cite{Rokhinson2006,McEuen1990,Nockel1994,Zailer1994}. In order to separate the generic 0.7 anomaly experimentally from these local conductance resonances we apply the following criteria.
\begin{enumerate}[(i)]
\item \emph{Temperature dependence}. While impurity resonances in the linear conductance become sharper at lower temperatures, the 0.7 anomaly is most pronounced at elevated temperatures and gradually disappears towards lower temperatures.
\item \emph{Zero bias anomaly (ZBA)}. As the 0.7 anomaly in the linear conductance vanishes with decreasing temperature, an anomalous peak around zero source-drain bias in the finite bias conductance spectrum emerges simultaneously. Resonances in the transmission due to impurities may also lead to a peak in the differential conductance at zero bias, however, the emergence of such a peak is not accompanied by a vanishing feature in the linear conductance.
\item \emph{Lateral electric fields}. The application of a transverse in-plane electric field results in a lateral shift of the QPC axis. It is expected to change any electronic coupling between the propagating and the impurity state, if present, leading to a corresponding change in the resonant feature. In contrast, the 0.7 anomaly is stable under the application of a transverse in-plane electric field.
\end{enumerate}
Based on the literature and our experience with several hole QPCs having different geometries and being fabricated with different techniques, we regard (i) and (ii) as \emph{conclusive evidences} while (iii) is considered as \emph{supportive evidence} for the unambiguous identification of the 0.7 anomaly. Note that the effect of a transverse lateral electric field on the 0.7 feature may depend on the QPC design. However, the 0.7 feature is expected to be unaffected in QPCs with symmetrical lateral confinement.

\begin{figure}[t!]
\includegraphics[width=\columnwidth]{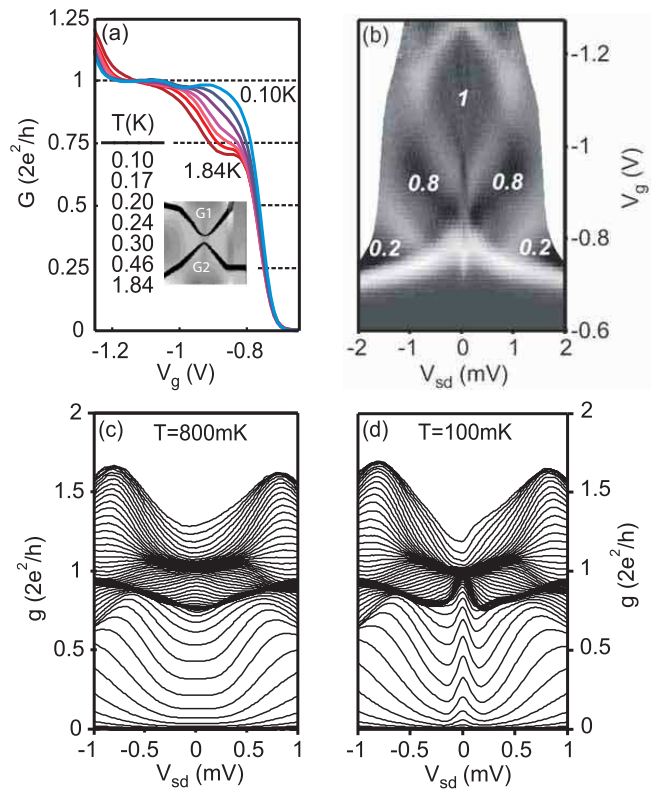}
\caption{(color online). (a) Linear conductance $G$ as a function of the gate voltage $V_{\rm g}$ at $B$ = 0 at temperatures between $T$ = 1.84 K and 100 mK. At elevated temperatures the emergence of an extra plateau at $\sim$0.7(2$e^{2}/h$) is clearly visible. The inset shows the micrograph of the QPC. The 2DHG is depleted under the etched black regions thus $G1$ and $G2$ can be used as in-plane gates to tune the electrical confinement of the QPC. The lithographical width of the QPC is 200 nm. (b) Transconductance [$dg(V_{\rm sd},V_{\rm g})/dV_{\rm g}$] in arbitrary units as a function of $V_{\rm sd}$ and $V_{\rm g}$ at $B$ = 0 and $T$ = 100 mK. Dark areas correspond to plateaus in $g$. The numbers indicate the conductivity values at the plateaus in units of 2$e^{2}/h$. (c)-(d) Non-linear differential conductance $g$ at $B$ = 0 as a function of $V_{\rm sd}$ taken at $T$ = 800 mK and 100 mK, respectively. Each trace correspond to different $V_{\rm g}$ gate voltages. Plateaus in $G$ appear as accumulation of the individual curves. The extra plateau around $V_{\rm sd}$ = 0 with $g\approx$ 0.7(2$e^{2}/h$) is only present at $T$ = 800 mK. With decreasing temperature a gradually emerging zero-bias anomaly (ZBA) restores the conductance at $2e^{2}/h$ around $V_{\rm sd}$ = 0.}
\label{zerofield.fig}
\end{figure}

The data presented here were acquired on a representative structure exhibiting a strong 0.7 anomaly in accordance with the above criteria. Similar data were obtained for more than 15 cool downs of 5 different samples having different designs and being fabricated with different technological methods. The sample presented here was patterned by electron beam lithography and wet chemical etching. The geometric confinement of the QPC is symmetric with respect to its axis, as displayed in the inset of Fig.~\ref{zerofield.fig}(a), and can be tuned by the G1 and G2 in-plane gates. The linear combination $V_{\rm g}=\alpha_{1}V_{\rm G1}+\alpha_{2}V_{\rm G2}$ tunes the confining potential in a symmetric fashion, where $-\alpha1/\alpha2$ is the slope of the lines of equal linear conductance on the $V_{\rm G1}$\,--\,$V_{\rm G2}$ plane. Accordingly, the asymmetric gate voltage combination $\Delta V=\alpha_{2}V_{\rm G1}-\alpha_{1}V_{\rm G2}$ leads to a transverse electric field resulting in a lateral shift of the QPC axis while the Fermi energy is kept constant by the leads.

Transport experiments were carried out between 4.2 K and the 100 mK base temperature of a $^{3}$He/$^{4}$He dilution refrigerator at magnetic fields of up to $B$ = 13 T applied perpendicular to the plane of the 2DHG. Measurements of the finite-bias differential conductance $g=dI(V_{\rm sd},V_{\rm g})/dV_{\rm sd}$ were carried out by the simultaneous application of an ac excitation with an amplitude of 20 $\mu$V at 31 Hz lock-in frequency, and a dc offset $V_{\rm bias}$ of up to $\pm$6 mV between source and drain. Four-terminal lock-in measurements of the linear conductance $[G=g(V_{\rm sd}=0)]$ were performed at a frequency of 31 Hz. The voltage drop $V_{\rm sd}$ across the QPC was measured using two independent leads. The conductance of the two-dimensional leads as a function of the magnetic field was simultaneously monitored by means of a two-terminal lock-in measurement via the two drain contacts of the QPC at a different frequency of 41 Hz. During data evaluation a magnetic field dependent conductance contribution arising from Shubnikov-de Haas conductance oscillations of the two-dimensional leads was carefully analyzed and separated from the conductance of the one-dimensional QPC channel. The field dependence of the subtracted serial resistance was in a good agreement with the one of the directly measured resistance of the drain lead justifying the procedure.

The standard zero magnetic field signatures of the 0.7 anomaly are demonstrated in Fig.~\ref{zerofield.fig}. At $T$ = 1.84 K the linear conductance $G$ exhibits a pronounced plateau-like feature at $0.7(2e^{2}/h)$. With decreasing temperature this extra feature gradually approaches the $(2e^{2}/h)$ first plateau until it completely disappears at $T$ = 100 mK, as shown in Fig.~\ref{zerofield.fig}(a). The comparison of the non-linear differential conductance recorded at $T$ = 800 mK and 100 mK shown in Fig.~\ref{zerofield.fig}(c)-(d) highlights the intimate relation of the 0.7 anomaly observed in $G$ to the ZBA, the narrow peak arising in $g$ around $V_{\rm sd}$ = 0 at low temperature. In Fig.~\ref{zerofield.fig}(c)-(d) the plateaus in $G(V_{\rm g})$ appear as the accumulation of the individual $g(V_{\rm sd})$ traces corresponding to different $V_{\rm g}$ gate voltages. This representation emphasizes the role of the ZBA in the low temperature recovery of the $(2e^{2}/h)$ unitary conductance from the 0.7 plateau \cite{Cronenwett2002}. Figure~\ref{zerofield.fig}(b) displays the low temperature transconductance $dg(V_{\rm sd},V_{\rm g})/dV_{\rm g}$ as obtained from Fig.~\ref{zerofield.fig}(d) by numerical differentiation. Dark regions of the gray scale map correspond to plateaus in $g$ with the plateau values indicated in units of $2e^{2}/h$. The bright areas represent the transitions between adjacent plateaus. Note that the cusp-like feature at the crossover between the finite bias 0.8 plateaus and pinch-off is the manifestation of the ZBA. It is to be emphasized that the data presented in Fig.~\ref{zerofield.fig} is very similar to n-type data reported in ~\cite{Cronenwett2002} testifying to the structural and electronic quality of our samples.

\begin{figure}[t!]
\includegraphics[width=\columnwidth]{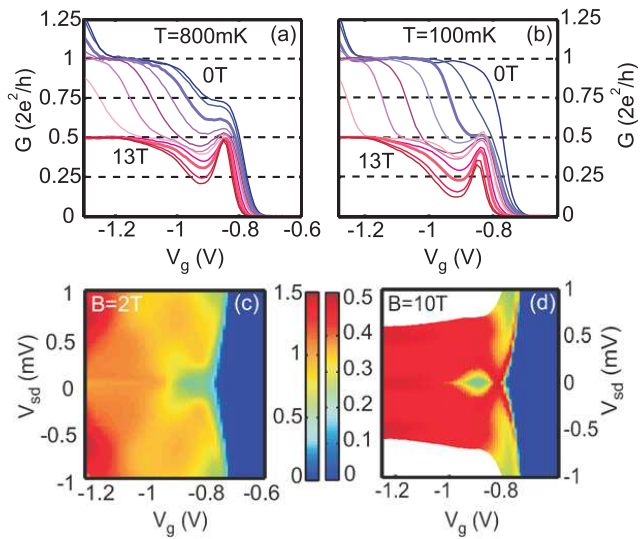}
\caption{(color online). (a)-(b) Linear conductance $G$ as a function of $V_{\rm g}$ at perpendicular magnetic fields ranging from $B$ = 0 T to 13 T at $T$ = 800mK and at $T$ = 100mK, respectively. (c)-(d) Finite bias differential conductance $g$ at $T$ = 100 mK shown for $B$ = 2 T [third trace from the top on panel (b)] and 10 T [third trace from the bottom on panel (b)], respectively. Note the different color scale for the left and right panels which are chosen to emphasize the diamond-like feature on the rise of the $0.5(2e^{2}/h)$ plateau.}
\label{Bdependence.fig}
\end{figure}

Figure~\ref{Bdependence.fig} shows the effect of a magnetic field applied \emph{perpendicular} to the plane of the 2DHG on the linear as well as on the finite-bias differential conductance. Studies of the 0.7 anomaly in this particular magnetic field orientation are rarely reported \cite{Gianetta2005}. Here we employ magnetic fields up to $B$ = 13 T in order to exploit localization phenomena which are potentially linked to the 0.7 anomaly. Figure.~\ref{Bdependence.fig}(a)-(b) exhibit the magnetic field dependence of $G(V_{\rm g})$ at $T$ = 800 mK and 100 mK, respectively. At $T$ = 800 mK the anomalous $0.7(2e^{2}/h)$ plateau gradually transforms into a pronounced peak which develops on the rise of the spin-split $0.5(2e^{2}/h)$ plateau at high magnetic field. At $T$ = 100 mK, although the 0.7 anomaly is lifted to the first plateau at zero field, the same qualitative behavior is observed as $B$ increases to 13 T.

The finite bias differential conductance spectra taken at $T$ = 100 mK are displayed for the representative magnetic field values $B$ = 2 T and 10 T in Fig.~\ref{Bdependence.fig}(c)-(d), respectively. They reveal that the formation of the diamond-like feature with reduced conductance in $g$ is associated with the peak that develops in $G$ as a function of the magnetic field [Fig.~\ref{Bdependence.fig}(a)-(b)]. It is reminiscent of the Coulomb blockade effect seen in quantum dots.

Considering similar experimental findings of Cronenwett \textit{et al.} ~\cite{Cronenwett2002} for electron systems along with the SDFT calculations \cite{Rejec2006} one can extend the model of the 0.7 anomaly based on the formation of a quasi-localized state in the QPC to the limit of high perpendicular magnetic field. The effect of this field is three-fold. (i) It acts on the spin degree of freedom via the energy separation of the spin subbands which is proportional to $B$ while (ii) the orbital part of the resonant wavefunctions shrinks as the magnetic length decreases with $\propto1/\sqrt{B}$. Additionally, (iii) the amplitude of the Friedel oscillations in the hole density around the bare QPC potential, which is believed to be responsible for the formation of the quasi-localized state \cite{Meir2008,Rejec2006}, is expected to be largely enhanced in high perpendicular magnetic fields \cite{Rensink1968}. At large magnetic fields the wavelength distribution of the holes is reduced to the magnetic length and its multiples depending on the number of the occupied Landau levels. Holes can contribute to screening only at these specific wavelengths leading to larger oscillations \emph{independently} of the dimensionality. As a result of (ii) and (iii), the coupling of the quasi-localized state to the source and drain reservoirs is expected to decrease and the localization of holes in the lower lying spin subband becomes more prominent, in agreement with the high temperature data shown in Fig.~\ref{Bdependence.fig}(a). We point out that as the amplitude of the Friedel oscillations is proportional to the effective mass of the screening carriers, stronger localization is expected for holes with an 8 times larger effective mass compared to electrons.

The observation that the 0.7 feature evolves smoothly into the resonance peak attributed to a Coulomb blockade-like effect suggests that a (quasi-)localized state of the same origin is involved in both phenomena as its coupling strengths to the two-dimensional leads gradually decreases with increasing magnetic field. Since the localization depends strongly on the strength of the confining barrier even small changes in that potential arising from stronger magnetic localization may have large effects on localization leading to a different qualitative nature of the localization at zero field and at the high perpendicular field limit. While the weak spin-dependent density oscillations described in \cite{Rejec2006} lead to the formation of a quasi-localized state which reduces the transmission of one of the two co-existing spin subbands, the Coulomb blockade effect at high perpendicular magnetic field reveals the existence of a resonant bound state of the fully spin-polarized holes. A speculation could be that the latter may occur when with increasing density half of the Fermi-wavelength matches the QPC length between the enhanced barriers and resonant tunneling can take place between the reflected edge states of the source and drain leads.

At $T$ = 100 mK [Fig.~\ref{Bdependence.fig}(b)] the 0.7 feature is suppressed and the $2e^{2}/h$ plateau is fully restored in $G$, in agreement with the Kondo model of the 0.7 anomaly where the spin of the quasi-localized carrier is expected to be dynamically screened by a Kondo-correlated collective state at zero magnetic field \cite{Cronenwett2002}. With increasing magnetic field orbital effects enhance the localization in the same manner but even more efficiently than at $T$ = 800 mK. This is reflected in the reduced differential conductance as well as in the sharper edges of the diamond in Fig.~\ref{Bdependence.fig}(d).

\begin{figure}[t!]
\includegraphics[width=\columnwidth]{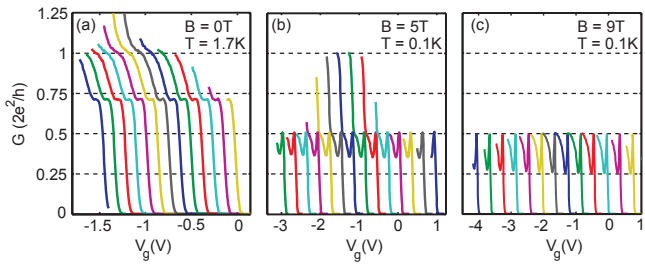}
\caption{(color online). (a)-(c) Linear conductance $G$ as a function of $V_{\rm g}$ at selected magnetic fields of $B$ = 0 T, $B$ = 5 T and $B$ = 9 T, respectively. The temperature is $T$ = 1.7 K for $B$ = 0 T and $T$ = 100 mK for the finite field data. Each conductance curve corresponds to a different $\Delta V$ value ranging from $\Delta V$ = -0.35 V (leftmost trace) to $\Delta V$ = 0.37 V (rightmost trace) on panel (a), from $\Delta V$ = -0.35 V (leftmost trace) to $\Delta V$ = 0.3 V (rightmost trace) on panel (b) and from $\Delta V$ = -0.4 V (leftmost trace) to $\Delta V$ = 0.2 V (rightmost trace) on panel (c). The individual traces are horizontally shifted for clarity. Note that the accessible conductance regime becomes more limited at high magnetic fields due to leakage to the side gates.}
\label{deltav.fig}
\end{figure}

The range of the applied perpendicular magnetic field covers different transport regimes in the leads from that of Shubnikov-de Haas oscillations to that of the integer quantum Hall states with edge channel conductivity. The effect of the Shubnikov-de Haas oscillations arising at intermediate magnetic fields is taken into account by subtracting a separately measured field dependent serial resistance from the total resistance. In the quantum Hall regime the transmission through the QPC can display unusual additional plateaus related to tunneling between edge states \cite{Haug1988}. However, one expects such resonances to appear at certain filling factor configurations in the QPC and the leads. The conductance resonance reported here has a smooth magnetic field dependence and is not filling factor dependent. In the magnetic field range 4-13 T the filling factor in the leads (QPC) varies from 4 to 1 (2 to 1).

In the absence of a localized state energy dependent resonant scattering between propagating edge channels of the leads and of the QPC \cite{Wees1991} is not expected to lead to a Coulomb diamond in the finite bias conductance. On the other hand, resonances between propagating and localized edge states encircling possible imperfections of the background potential inside the QPC \cite{Jain1988} may result in observations similar to those displayed in Fig.~\ref{Bdependence.fig}(c)-(d). However, this issue can be addressed experimentally by applying transverse electric fields to shift the edge channels with respect to the localized states.

The generic nature of the effect is therefore further supported by experiments performed under different asymmetric in-plane gate voltages. Nonzero $\Delta V$ values result in a lateral shift of the QPC channel within the plane of the 2DHG. Changing $\Delta V$ has a negligible effect on the conductance traces as shown in Fig.~\ref{deltav.fig} for zero field as well as for magnetic fields corresponding to Landau level filling factors of 3 and 2 in the two-dimensional leads at $B$ = 5 T [Fig.~\ref{deltav.fig}(b)] and $B$ = 9 T [Fig.~\ref{deltav.fig}(c)], respectively. The lateral shift of the QPC axis as a function of $\Delta V$ is estimated to be 8-10 nm/V \cite{Heinzel2000}. As the Bohr radius in p-GaAs is in the order of 1-2 nm this covers a sufficiently large region to exclude impurity resonance effects. The smooth transition of the 0.7$(2e^{2}/h)$ plateau into a Coulomb resonance peak with increasing magnetic field has been found to be robust also against thermal cycling.

In conclusion, we have provided experimental evidence for the importance of a quasi-localized state and Coulomb blockade physics for the 0.7 anomaly. By taking advantage of the enhanced screening properties of low-dimensional holes compared to electrons we have shown that at high perpendicular magnetic fields the coupling of the quasi-localized state to the leads gradually decreases and, as a consequence, the 0.7$(2e^{2}/h)$ plateau evolves smoothly into a robust resonance peak residing at the rise of the 0.5$(2e^{2}/h)$ plateau due to enhanced Coulomb blockade. The generic origin of the quasi-localized state has been demonstrated by the application of transverse in-plane electric fields.

Helpful discussions with C. R\"ossler and U. Gasser are acknowledged. This research was supported by the Swiss National Science Foundation, the German Science Foundation and the German Ministry for Science and Education. A.D.W. and D.R. thank the SFB491, SPP1285 and the BMBF nanoQUIT for financial support. Y.M. has been supported by the ISF. M.C. is grateful to the European Commission for financial support under a Marie Curie Intra European Fellowship within the 7$^{\rm th}$ European Community Framework Programme.


\begin{references}

\bibitem{Wees1988} B. J. van Wees \textit{et al.}, Phys. Rev. Lett. {\bf60}, 848 (1988).

\bibitem{Wharam1988} D. A. Wharam \textit{et al.}, J. Phys. C {\bf21}, L209 (1988).

\bibitem{Buttiker1990} M. B\"uttiker, Phys. Rev. B {\bf41}, 7906 (1990).

\bibitem{Thomas1996} K. J. Thomas \textit{et al.}, Phys. Rev. Lett. {\bf77}, 135 (1996).

\bibitem{Kristensen2000} A. Kristensen \textit{et al.}, Phys. Rev. B {\bf62}, 10950 (2000).

\bibitem{Berggren2002} K. -F. Berggren and I. I. Yakimenko, Phys. Rev. B {\bf66}, 085323 (2002).

\bibitem{Reilly2002} D. J. Reilly \textit{et al.}, Phys. Rev. Lett. {\bf89}, 246801 (2002).

\bibitem{Reilly2005} D. J. Reilly, Phys. Rev. B {\bf72}, 033309 (2005).

\bibitem{Rokhinson2006} L. P. Rokhinson, L. N. Pfeiffer, and K. W. West, Phys. Rev. Lett. {\bf96}, 156602 (2006).

\bibitem{Sfigakis2008} F. Sfigakis \textit{et al.}, Phys. Rev. Lett. {\bf100}, 026807 (2008).

\bibitem{Rejec2000} T. Rejec, A. Ram\v{s}ak, and J. H. Jefferson, Phys. Rev. B {\bf62}, 12985 (2000).

\bibitem{Matveev2004} K. A. Matveev, Phys. Rev. B {\bf70}, 245319 (2004).

\bibitem{Meir2002} Y. Meir, K. Hirose, and N. S. Wingreen, Phys. Rev. Lett. {\bf89}, 196802 (2002).

\bibitem{Meir2008} Y. Meir, J. Phys.: Condens. Matter {\bf20}, 164208 (2008).

\bibitem{Hirose2003} K. Hirose, Y. Meir, and N. S. Wingreen, Phys. Rev. Lett. {\bf90}, 026804 (2003).

\bibitem{Rejec2006} T. Rejec and Y. Meir, Nature {\bf442}, 900 (2006).

\bibitem{Guclu2009} A. D. G\"u\c{c}l\"u \textit{et al.}, Phys. Rev. B {\bf 80}, 201302(R) (2009).

\bibitem{Cronenwett2002} S. M. Cronenwett \textit{et al.}, Phys. Rev. Lett. {\bf88}, 226805 (2002).

\bibitem{Grbic2007} B. Grbi\'{c} \textit{et al.}, Phys. Rev. Lett. {\bf99}, 176803 (2007).

\bibitem{Komijani2008} Y. Komijani \textit{et al.}, Europhys. Lett. {\bf84}, 57004 (2008).

\bibitem{Hamilton2008} A. R. Hamilton \textit{et al.}, J. Phys.: Condens. Matter {\bf20}, 164205 (2008).

\bibitem{Danneau2008} R. Danneau \textit{et al.}, Phys. Rev. Lett. {\bf100}, 016403 (2008).

\bibitem{Klochan2006} O. Klochan \textit{et al.}, Appl. Phys. Lett. {\bf89}, 092105 (2006).

\bibitem{Shabani2008} J. Shabani \textit{et al.}, Appl. Phys. Lett. {\bf93}, 212101 (2008).

\bibitem{Reuter1999} D. Reuter, A. D. Wieck, and A. Fischer, Rev. Sci. Inst. {\bf70}, 3435 (1999).

\bibitem{Grbic2004} B. Grbi\'{c} \textit{et al.}, Appl. Phys. Lett. {\bf85}, 2277 (2004).

\bibitem{McEuen1990} P. L. McEuen \textit{et al.}, Surf. Sci. {\bf229}, 312 (1990).

\bibitem{Nockel1994} J. U. N\"ockel and D. Stone, Phys. Rev. B {\bf50}, 17415 (1994).

\bibitem{Zailer1994} I. Zailer \textit{et al.}, Phys. Rev. B {\bf49}, 5101 (1994).

\bibitem{Gianetta2005} R. W. Giannetta \textit{et al.}, Physica E {\bf27}, 270 (2005).

\bibitem{Rensink1968} M. E. Rensink, Phys. Rev. {\bf174}, 744 (1968).

\bibitem{Haug1988} R. J. Haug \textit{et al.}, Phys. Rev. Lett. {\bf61}, 2797 (1988).

\bibitem{Wees1991} B. J. van Wees \textit{et al.}, Phys. Rev B {\bf43}, 12431 (1991).

\bibitem{Jain1988} J. K. Jain and S. A. Kivelson, Phys. Rev. Lett. {\bf60}, 1542 (1988).

\bibitem{Heinzel2000} T. Heinzel \textit{et al.}, Phys. Rev. B {\bf61}, R13353 (2000).

\end{references}
\end{document}